\def\CA{\hbox{{$\cal A$}}}

\def\CT{\hbox{{$\cal T$}}}
\def\o{{}_{\scriptscriptstyle(1)}}
\def\t{{}_{\scriptscriptstyle(2)}}
\def\ba{{\bf A}}
\def\<{\langle}
\def\>{\rangle}
\def\eps{{\epsilon}}
\def\tens{\mathop{\otimes}}
\def\la{{\triangleright}}

\def\id{\hbox{id}}
\def\beq{\begin{equation}}
\def\eeq{\end{equation}}
\def\bea{\begin{eqnarray}}
\def\eea{\end{eqnarray}}

\def\={\; = \;}

\def\bt{\bullet}

\def\DGL{{{}_\Gamma}\Delta}
\def\DGR{\Delta_\Gamma}
\def\D{\Delta}
\def\om{\omega}
\def\id{\hbox{id}}
\def\mk{{{\cal M}^4_\kappa}}
\def\kp{\frac{1}{\kappa}}
\def\bt{\bullet}

\documentstyle[12pt]{article}

\begin{document}
\begin{titlepage}
\begin{flushright}
DAMPT 94/?? \\
hep-th/9409014
\end{flushright}

\vspace{0.5in}
\setcounter{footnote}{1}
\def\thefootnote{\fnsymbol{footnote}}
\begin{center}
{\LARGE NONCOMMUTATIVE DIFFERENTIAL CALCULUS \\
\ \\
ON THE $\kappa$-MINKOWSKI SPACE}\\
\ \\
{\ }\\ Andrzej Sitarz
\footnote{Partially supported by KBN Grant 2P 302 103 06}
\footnote{Permanent address: Department of Field Theory,
 Institute of Physics,
Jagiellonian University, Reymonta 4, 30-059 Krak\'ow, Poland,
{\it e-mail: sitarz@if.uj.edu.pl} }
\def\thefootnote{\arabic{footnote}}

\vskip.5in

Department of Applied Mathematics \& Theoretical Physics\\ University of
Cambridge, Cambridge CB3 9EW, U.K.\\

\vskip.5in

September 1994
\end{center}

\vskip 0.5in

\begin{quote}
\noindent{\bf ABSTRACT\ }

Following the construction of the $\kappa$-Minkowski space from
the bicrossproduct structure of the $\kappa$-Poincare group, we
investigate possible differential calculi on this noncommutative
space. We discuss then the action of the Lorentz quantum algebra
and prove that there are no 4D bicovariant differential calculi,
which are Lorentz covariant. We show, however, that there exist a
five-dimensional differential calculus, which satisfies both
requirements. We study also a toy example of 2D $\kappa$-Minkowski
space and and we briefly discuss the main properties of its
differential calculi.

\end{quote}
\end{titlepage}
\setcounter{footnote}{0}
\section{The $\kappa$-Poincare algebra}

The $\kappa$-Poincare algebra has been introduced \cite{LNRT,LNR}
and studied extensively \cite{N,LRR,B,D,R} as one of possible Hopf algebra
deformations of the standard Poincare algebra.
The momenta and the rotation generators remain unchanged and
the deformation occurs only in the boost sector and the coproduct
structure.

Recently, it was shown \cite{MR} that the $\kappa$-Poincare has a structure
of a Hopf algebra extension of (classical) $U(so(1,3))$ by the Hopf algebra of
(deformed) translations $\CT$:
\bea
&[P_{\mu},P_{\nu}] = 0,& \label{p1} \\
&\Delta P_0=P_0 \tens 1 + 1 \tens P_0,& \label{p2} \\
&\Delta P_i=P_i \tens 1 + e^{-{P_0\over \kappa}} \tens P_i.& \label{p3}
\eea

The commutation relations of the Lorentz algebra generators,
rotations $M_i$ and boosts $N_i$ are the standard ones:
\beq
[M_i,M_j]= \eps_{ijk}M_k, \quad
[M_i,N_j] =\eps_{ijk}N_k, \quad
[N_i,N_j]= - \eps_{ijk} M_k,
\eeq
whereas the cross relations and the coproduct structure of the Lorentz
part are deformed:
\bea
& [P_0,M_i] = 0, \quad [P_i,M_j] = \eps_{ijk} P_k & \\
& [P_0,N_i] = -P_i, \quad [P_i,N_j]=-\delta_{ij}
( {\kappa \over 2}(1-e^{-{2 P_0\over\kappa}}) +
{1\over 2\kappa}{\vec P}^2 ) + {1 \over \kappa} P_i P_j, &\\
& \Delta N_i = N_i \tens 1+ e^{-{P_0\over\kappa}}\tens
N_i+{\eps_{ijk}\over \kappa} P_j \tens M_k,\quad
\Delta M_i=M_i\tens1+1\tens M_i. &
\eea
The classical Poincare algebra is obtained in the limit
$\kappa \to \infty$.

\section{$\kappa$-Minkowski space}

Here we shall briefly outline the construction of the
$\kappa$-Minkowski space  ($\mk$) and the action of the $\kappa$-Poincare
on its generators, as developed in \cite{MR}.

As the $\kappa$ deformation of Minkowski space we take the dual
Hopf algebra of the translation algebra $\CT$ and we denote its
generators by $x_\mu$. From the relations (\ref{p1}-\ref{p3}) we
immediately obtain:
\bea
&[x_i,x_j]=0, \quad
[x_i,x_0] = {x_i\over\kappa}, & \\
&\Delta x_\mu=x_\mu\tens 1+1\tens x_\mu.& \label{x-co}
\eea
The canonical action of translations on our Minkowski space is:
\beq
t \la x = <x\o,t> x\t,\quad \forall x\in \CT^*, \quad t\in \CT,
\eeq
where we use shorthand notation for $\Delta x = \sum x\o \tens x\t$.

{}From the bicrossproduct structure of $\kappa$-Poincare we have the
action of $U(so(1,3))$ on translations $\CT$, which now, by duality,
could be translated into action on the generators of the Minkowski space:
\beq
M_i\la x_j=\eps_{ijk}x_k, \quad M_i\la x_0=0,\quad
N_i\la x_j=-\delta_{ij}x_0, \quad N_i\la x_0=-x_i, \label{Lac}
\eeq
which generalizes to the whole algebra by the covariance condition:
\beq
h \la xy = (h\o \la x) (h\t \la y), \quad \forall h\in U(so(1,3)),
x,y \in T^*, \label{Lac-co}
\eeq
Finally, let us quote here another result of \cite{MR} that the
lowest order Lorentz-invariant polynomial in $x_\mu$ is:
\beq
x^2_0-{\vec x}^2+{3\over \kappa}x_0. \label{metric}
\eeq

\section{Differential calculus}

In this section we shall present the main points of the construction
of bicovariant differential calculi on the $\kappa$-Minkowski space.
Let us remind that the (first order) bicovariant calculus on a Hopf
algebra ${\CA}$ is defined by $(\Gamma, \DGL, \DGR,d)$, where
$\Gamma, \DGR,\DGL$ defines a bicovariant bimodule over ${\CA}$
(for details see \cite{WOR}). In particular:
$ \DGL: \Gamma \to \Gamma \tens \CA$,
$ \DGR : \Gamma \to \CA \tens \Gamma$, are such that:
\bea
& \DGR(a\om) = \D(a)\DGR(\om), \;\;\; \DGR(\om a) =
 \DGR(\om)\D(a), \nonumber \\
& \DGL(a\om) = \D(a)\DGL(\om), \;\;\; \DGL(\om a) =
 \DGL(\om)\D(a), \nonumber \\
& (\D \tens \id)\DGR = (\id \tens \DGR)\DGR, \nonumber \\
& (\id \tens \D)\DGL = (\DGL \tens \id)\DGL,  \nonumber \\
& (\id \tens \DGL)\DGR = (\DGR \tens \id)\DGL.  \nonumber
\eea
In addition, the external derivative $d$ satisfies:
\beq
 \DGL d = (d \tens \id) \D, \;\;\; \DGR d = (\id \tens d) \D.
 \label{biv_d}
\eeq
Now, we shall investigate the case of bicovariant differential
calculi on the $\kappa$-Minkowski space. Let us begin with
observation, that, due to the coproduct structure on $\mk$ (\ref{x-co}) and
the bicovariance property (\ref{biv_d}) all $dx_\mu$ are
simultaneously left and righ-covariant:
\beq
\DGR(dx_\mu) = 1 \tens dx_\mu, \;\;\; \DGL(dx_\mu) = dx_\mu \tens 1.
\label{bicov_1}
\eeq
Let us denote the basis of left-invariant forms as $\chi_a$,
$a=0,\ldots,N$, $N \geq 4$.
By simple calculation we verify that the commutator $[x_\mu, \chi_a]$
is also left-invariant:
\beq
\DGR ( [x_\mu, \chi_a] ) = 1 \tens  [x_\mu, \chi_a],
\eeq
therefore, from the general properties of the bicovariant bimodules
(Theorem 2.1 \cite{WOR} ) we deduce that the commutator $[x_\mu. \chi_a]$
must have the following expansion:
\beq
[x_\mu, \chi_a] = \sum_{\mu,a,b} A_{\mu a}^b \chi_b.
\label{x-dx-rel}
\eeq
There are, of course, some consistency conditions for the
above relations, which come from the mixed Jacobi identity:
\beq
[ [x_\mu , x_\nu] , \chi_a ] + [ [x_\nu ,\chi_a ], x_\mu ] +
[ [ \chi_a , x_\mu ], x_\mu ] = 0. \label{jacobi}
\eeq
If we rewrite, for simplicity  of notation, the commutation relations
(\ref{x-co}) on $\mk$, in a more general form:
\beq
 [x_\mu, x_\nu] = B_{\mu\nu}^\rho x_\rho,
\label{x-co2}
\eeq
the relations (\ref{jacobi}) may be rewritten as:
\beq
 A_{\nu c}^a A_{\mu b}^c - A_{\mu c}^a A_{\nu b}^c
 = B_{\mu\nu}^\rho A_{\rho b}^a,
\label{cond_1}
\eeq
which simply state that the map $\pi: x_\mu \to -\ba_\mu$, where
$\ba_\mu$ is an $N \times N$ matrix, is a representation of the Lie
algebra defined by the relations (\ref{x-co}). Therefore the general
theory of bicovariant bimodules on the $\kappa$-Minkowski space is
linked with the finite dimensional representation theory of the
Lie algebra generated by $x_\mu$.
\footnote{Of course, this results are general and may be applied to
construction and classification of bicovariant bimodules and bicovariant
differential calculi on any universal enveloping algebra. In this paper,
however, we restrict ourselves only to the case of $\kappa$ Minkowski
space.}

Furthermore, due to (\ref{bicov_1}) we obtain that $dx_\mu$ may be
expressed as a linear combination of~$\chi_a$:
\beq
dx_\mu = D_\mu^a \chi_a,
\eeq
and if we impose the Leibnitz rule, by differentiating (\ref{x-co2})
we obtain another restriction:
\beq
D_\nu^b A_{\mu b}^a - D_\mu^b A_{\nu b}^a = B_{\mu\nu}^\rho D_\rho^a.
\label{cond_2}
\eeq
Both relations (\ref{cond_1},\ref{cond_2}) are sufficient and necessary
to determine a bicovariant differential calculus on the $\kappa$-Minkowski
space. In what follows, we shall not attempt, however, to classify all
possible bicovariant differential calculi, as we shall exploit
other restrictions provided by rich structure of the $\kappa$-Poincare
algebra.

\section{The action of $\kappa$-Lorentz on the differential calculus.}

Having discussed the structure of the possible bicovariant differential
calculi on the  $\kappa$-Minkowski space we shall now proceed to extend
the action of the Lorentz algebra to the bimodule of one-forms.

We shall postulate that the action of the Lorentz algebra
(\ref{Lac}-\ref{Lac-co}) extends to the differential algebra
in a natural covariant way, i.e.:\footnote{In fact, one may
postulate as well that the action of translations (as discussed
in \cite{MR}) extends in the same way, however, as
as $P_\mu \la dx_\nu = 0$,  due to (\ref{cov}) and the
coproduct structure (\ref{p3}) one may see that every bicovariant
differenctial calculus with relations (\ref{x-co2}) is
automatically covariant in the above sense with respect to the
action of translations alone. Therefore, we shall concentrate
on the highly nontrivial Lorentz part of the action.}

\beq
h \la (y \, dx) = (h\o \la y) \left( d( h\t \la x) \right){,} \quad
h \la (dx\, y) = \left( d (h\o \la x) \right) (h\t \la y). \label{cov}
\eeq

{}From the above definition and the action (\ref{Lac}) we
easily obtain the following identities:
\bea
& N_k \la [x_i, dx_j] = -\delta_{ki} [ x_0, dx_j ] - \delta_{kj}
[ x_i, dx_0 ] + {\kp} \left( \delta_{kj} dx_i
- \delta_{ij} dx_k \right),
\label{l_1} \\
& N_k \la [x_0, dx_i] = - [x_k,dx_i ] - \delta_{ki} [x_0, dx_0 ]
+ {\kp} \delta_{ki} dx_0,
\label{l_2} \\
& N_k \la [x_i, dx_0] = - [x_i,dx_k ] - \delta_{ki} [x_0, dx_0 ],
\label{l_3} \\
& N_k \la [x_0, dx_0] = - [x_k,dx_0 ] -  [x_0, dx_k ]
+ {\kp} dx_k,
\label{l_4} \\
& M_k \la [x_i, dx_j] = \epsilon_{kis} [x_s, dx_j]
+ \epsilon_{kjs} [x_i, dx_s],
\label{l_5} \\
& M_k \la [x_0, dx_i] = \epsilon_{kis} [x_0, dx_s]
\label{l_6} \\
& M_k \la [x_i, dx_0] = \epsilon_{kis} [x_s, dx_0]
\label{l_7} \\
& M_k \la [x_0, dx_0] = 0.
\label{l_8}
\eea

{}From the above relations we can immediately see that if we
postulate the $\kappa$-Lorentz covariance we can no longer
have a commutative differential calculus on the subalgebra
generated by $x_i$, $i=1,2,3$, as we must have (at least)
a non-vanishing term:
\beq
[x_i, dx_j] \; = \; \delta_{ij} {\kp} dx_0 + \ldots,
\eeq
which follows directly from the equation (\ref{l_1}).

Now, if we consider only 4D bicovariant calculi, with the bimodule
of one forms generated by $dx_\mu$, taking into account the
relations (\ref{x-dx-rel}) (with $\chi_\mu=dx_\mu$) and plugging
them into (\ref{l_1}-\ref{l_8}), we obtain system of linear
equations for the coefficients $A_{\mu\nu}^\rho$, which we
can solve. It appears, that the solution is unique and
gives us the following relations:
\bea
&[x_i, dx_j] =  \delta_{ij} {\kp} dx_0, \quad [x_i, dx_0]
= {\kp} dx_i, \\
&[x_0, dx_i] = 0, \quad [x_0, dx_0] = 0,
\eea
which, however, do not  define a differential calculus as they
fail to obey the condition (\ref{cond_1})! Therefore we may
conclude that there exist no bicovariant and $\kappa$-Lorentz
covariant 4D differential calculus on the
$\kappa$-Minkowski space. Before we proceed with further
considerations of the 4D situation, let us study a much simpler
model, of 2D $\kappa$-Poincare and $\kappa$-Minkowski space.

\section{2D $\kappa$-Minkowski space and the differential calculi}

The two-dimensional $\kappa$-Poincare algebra is defined in
a similar way as the 4D one (see \cite{LR2} for details.) For simplicity
of notation let us denote the momentum generators as $E$ and $P$
and the boost operator by $N$. Then the commutation relations and
the coproduct structure are as follows:
\bea
& [P,E] = 0 \;\;\; [N,E]=P, \;\;\;
[N,P] = - \frac{\kappa}{2}(1-e^{-{2E\over \kappa}}) + \frac{1}{2\kappa} P^2, \\
& \D P = P \tens 1 + e^{-{E\over \kappa}} \tens P, \;\;\;
\D E = E \tens 1 + 1 \tens E, \;\;\;
\D N =  N \tens 1 + e^{-{E\over \kappa}} \tens N.
\eea
The 2D $\kappa$-Minkowski could be defined like in the 4D case as
a Hopf dual to the algebra of momenta. If we call its generators $x,t$,
we shall have:
\beq
[x , t] = {\kp} x, \;\;\;
\D x = x \tens 1 + 1 \tens x, \;\;\;
\D t = t \tens 1 + 1 \tens t.
\eeq
The bicrossproduct construction could be also repeated in this case,
and the action of the boost $N$, which follows from it is:
\beq
N \la t = -x, \;\;\; N \la x = -t.
\eeq
which extends onto the whole of the algebra according to the
covariance prescription (\ref{Lac-co}).

\subsection{2D differential calculus}

Using the same arguments as in the 4D situation, we shall
postulate both bicovariance and Lorentz covariance of the
differential calculus. First let us present the relations
following from the latter requirement:
\bea
& N \la [t, dt] = - [x, dx] - [t, dx] + {\kp} dx, \label{2d-1} \\
& N \la [x, dt] = - [t, dt] - [x ,dx], \label{2d-2} \\
& N \la [t, dx] = - [t, dt] - [x, dx] + {\kp} dt, \label{2d-3} \\
& N \la [x, dx] = - [t, dx] - [x, dt], \label{2d-4}.
\eea
Again, if we look for two-dimensional bicovariant calculi, which satisfy
the above covariance property, we may solve the corresponding system of
linear equations, obtaining:
\beq
[x,dx]={\kp}dt, \;
[x,dt]={\kp}dx, \;
[t,dx]=0, \;
[t,dt]=0,
\eeq
which, though being the solution to (\ref{2d-1}-\ref{2d-4}),
does not give a differential calculus, since it fails to obey
(\ref{cond_1}). This result is hardly surprising, as we
already know that this was the case in four dimensions.
Therefore we must look for other possibilities, the simplest
of which, is the higher-dimensional calculus.

\subsection{3D differential calculus}

Let us assume that the bimodule of one-forms is generated by
left-invariant forms $dx,dt$ and $\phi$.
Furthermore, we shall assume that $N \la \phi = 0$.
{}From the general theory we already know (\ref{x-dx-rel}) the most
general form of the commutators. We may always choose $\phi$ in such
a way that $[x , \phi] = \alpha dx$~\footnote{By rescaling $\phi$
we may set $\alpha=1$ unless it is $0$, the latter case, however,
does not give any solutions.}

Then, by applying $N$ to both sides, we verify that the only
consistent case of Lorentz covariant calculus is:
\beq
[x ,\phi] = dx, \;\;\; [t, \phi] = dt + \phi.
\eeq
while the rest of the relations are still unknown:
\beq
[x_\mu, \chi_a] = A_{\mu a}^b \chi_b, \;\; t=x_0,  \; x=x_1;\;
\chi_0=dt,\; \chi_1=dx,\; \chi_2=\phi.
\eeq

Now, solving the system of linear equations (\ref{2d-1}-\ref{2d-4})
for the coefficients $A_{\mu a}^b$ restricts the number of free
parameters from $12$ to $2$!~\footnote{As the calculations are
straightforward, we shall not present them here.} If we denote the free
parameters as $a,b$, the matrices $\ba_t,\ba_x$ introduced earlier
to define the commutation relations are:
$$
\ba_t = \left(
\begin{array}{lll}
0 & 0 & a \\
0 & 0 & b \\
1 & \beta & 0 \\ \end{array}
\right), \;\;\;\; \ba_x = \left( \begin{array}{lll}
0 & {{\kp}} & -b \\
{\kp} & 0 & -a \\
0 & 1 & 0 \\ \end{array} \right){.} $$
Now, if we impose both consistency conditions (\ref{cond_1},\ref{cond_2})
it appears that they would be satisfied only and only if $a={\kp}$
and $b=0$.

Therefore, there exist only one 3D bicovariant and Lorentz covariant
differential calculus on the 2D $\kappa$-Minkowski space, with the
following commutation relations:

\bea
& [x,dx]={\kp}(dt- {\kp}\phi), \;\;
[x,dt]={\kp}dx, \;\;
[x,\phi]=dx, \nonumber \\
& [t,dx]=0, \;\;
[t,dt]=\frac{1}{\kappa^2}\phi, \;\;
[t,\phi]=dt. \label{2D-dif}
\eea

Before we turn back to the four-dimensional situation let us comment
briefly on the obtained result. First of all, let us rewrite slightly
the relations (\ref{2D-dif}) introducing an one-form
$\psi= dt - {\kp}\phi$:
\bea
& [x,dx]= \kp \psi, \;\; [x, \phi] = dx, \;\; [x, \psi] = 0, \nonumber \\
& [t,dx]= 0, \;\; [t, \psi] = - \kp \psi,\;\;
[t, dt] = \frac{1}{\kappa^2} \phi.
\label{2D-rewr}
\eea
Though it is merely a simply change of basis, we find it convenient to
present the rules of differentiation for elements of algebra
constructed of $x$ and $t$ alone and interpreted as usual
functions (polynomials) on the real line. In the case of $x$
we have:
\beq
d f(x) = dx\; \partial_x f(x) + \psi\; \frac{1}{2\kappa} \partial_{xx} f(x),
\eeq
and since the commutation relations between $x,dx,\psi$ are closed,
this defines a differential submodule of our bigger one. Let us
point out here that this specific $\kappa$-deformed calculus on
the real line obtained as a restriction of the bigger structure is
equivalent to the example of higher-derivatives calculus discussed
elsewhere \cite{SIT}.

For the $t$ variable the differential structure is, however, much
different. It is convenient to use the forms $\psi$ and
$\tilde{\psi}= dt+\kp \phi$, which have much simpler commutation
relations with $t$:
\beq
[t, \psi] = -\kp \psi, \;\;\;\; [t, \tilde{\psi} ] = \kp \tilde{\psi}.
\eeq
Now it appears that:
\beq
d f(t) = \tilde{\psi}\;  (1-e^{\kp\partial_t}) f(t) +
\psi\;  (1-e^{-\kp\partial_t}) f(t).
\eeq
This differential calculus is a system is far more complicated
than the one considered earlier, as it involves the partial
derivatives of all orders.

Finally, let us consider the action of the external derivative $d$
on an arbitrary function $f(t,x)$, with the normal ordering $:f:$,
which denotes that all powers of $t$ are shifted to the left:
\beq
df = dx \; \partial_x :f: + \psi\; (1 - e^{-\kp\partial_t} +
\frac{1}{2\kappa}  e^{-\kp\partial_t} \partial_{xx} ) :f:
+ \tilde{\psi}\;  (1 - e^{\kp\partial_t} ) :f:,
\eeq
which is a highly complicated expression.

Further, we may construct higher-order forms, and it appears
that the following set of rules defines the multiplication of
one-forms:
\bea
& dx \bt dt = - dt \bt dx, \;\; dt \bt \phi = - \phi \bt dt, \;\;
dx \bt \phi = - \phi \bt dx, \nonumber \\
& dx \bt dx = - dt \bt dt, \;\;\;\; d\phi = \kappa^2 (dt \bt dt - dx \bt dx).
\eea
This is the weakest set of constraints on the higher-order calculus,
we may as well consider the quotient of the above with $dx \bt dx =0$,
thus enforcing $dt \bt dt =0$ and $ d \phi = 0$.

\section{5D Lorentz covariant, bicovariant differential calculus
on 4D $\kappa$-Minkowski space}

Following the example of the last section with the toy model
of $2D$ bicovariant calculus, we shall attempt to construct
a corresponding structure in the four-dimensional situation.

Let us take the bimodule of one-forms generated by $dx_\mu$, $\mu=0,...,3$
and an additional one-form $\phi$. Moreover, motivated by the 2D
example, we shall assume that the form $\phi$ is Lorentz invariant:
\beq
N_i \la \phi = 0, \;\;\; M_i \la \phi = 0,
\eeq
and the commutation relations between the coordinates $x_\mu$ and all
generating one-forms $\phi$ are the following:
\bea
& [ x_\mu, \phi ] = dx_\mu, \;\;\; [x_0, dx_0] = \frac{1}{\kappa^2} \phi,
\nonumber \\
& [x_i, x_j ] = \delta_{ij} {\kp} (dx_0 -  {\kp} \phi),
\;\; [x_0, dx_i] = 0, \;\; [x_i, dx_0] =  {\kp} dx_i.
\eea
It could be easily checked that these relations satisfy the
Lorentz covariance conditions (\ref{l_1}-\ref{l_8}), and both
remaining conditions (\ref{cond_1}) and (\ref{cond_2}) and
therefore they define a five-dimensional bicovariant and Lorentz
covariant differential calculus on the $\kappa$-Minkowski space.

The differential calculus presented above has the same structure
as its two-dimensional analogue and one can repeat all steps
and calculate the explicit expression for $d$ and products
of higher order form. We shall not do it here, as the results
are exactly the same is in the 2D case, though, of course,
a single space variable $x$ should now be replaced by a triple
$x_1,x_2,x_3$.

We have demonstrated here the existence of such calculus,
however, we cannot yet claim that it is unique, though it seems
to be a reasonable hypothesis, and we shall address in future work.

\section{Conclusions}

The $\kappa$ deformation of the Poincare algebra is a good
example and a testing ground for possible deformation of
physical theories. The construction of the $\kappa$-Minkowski
space enables us to use the tools noncommutative geometry
to construct $\kappa$ deformations of field theory.
The differential calculus, being the most important tool,
is therefore a crucial point of these efforts. As we have
learned from the studies of quantum groups, the requirement
of bicovariance, though a strong one, seems to be at least
an elegant way of choosing a particular differential structure.
Of course, there may by many bicovariant differential calculi,
also in our case of the $\kappa$-Minkowski space.\footnote{Let
us mention here the simplest (and natural) example \cite{MR2}:
$[x_0, dx_i] = -{\kp} dx_i$, and other commutators vanishing.}
Therefore, one may look for further constraints, which in our
case are provided naturally by the action of the Lorentz algebra.
The requirement of Lorentz covariance (or rather full $\kappa$-Poincare
covariance, since, as we have already mentioned every bicovariant
calculus is automatically covariant with respect to the action of
translations)
covariance seems to be a reasonable choice, though its consequences
are much more significant than
one would suspect. The fact that there are no differential calculi
of the same dimension as the underlying space (which we have shown
for $D=2$ and $D=4$) is very interesting, though hardly surprising
in the $q$- or $\kappa$-deformed world. Let us stress, however, the
difference: here it is not only the bicovariance, which enforces it,
but some more requirements coming from some 'external' symmetries.

The higher-dimensional differential calculus, which we have constructed
in both cases is therefore the most reasonable candidate as a tool
for constructing models of $\kappa$-deformed field theory. This, as
well as some detailed studies of its properties, is an interesting
topic, which we shall investigate in future. What physical
consequences this calculus may have and how would contribute to
verification of $\kappa$-deformed physics as a feasible theory
is also an open problem.

\begin{flushleft}
{\large \bf Aknowledgements:}
\end{flushleft}

The author would like to thank Cambridge College Hospitality Scheme,
Sidney Sussex College and DAMPT (special thanks to R.Horgan)
for hospitality, and Shahn Majid for valuable remarks.

\end{document}